\documentclass[%
 aip,
 amsmath,amssymb,
 reprint,%
]{revtex4-1}

\usepackage{graphicx}
\usepackage{dcolumn}
\usepackage{bm}

\usepackage[utf8]{inputenc}
\usepackage[T1]{fontenc}
\usepackage{mathptmx}
\usepackage{etoolbox}

\usepackage{subfig}
\usepackage{comment}
\usepackage{siunitx}
\usepackage{textcomp,mathcomp}

\makeatletter
\def\@email#1#2{%
 \endgroup
 \patchcmd{\titleblock@produce}
  {\frontmatter@RRAPformat}
  {\frontmatter@RRAPformat{\produce@RRAP{*#1\href{mailto:#2}{#2}}}\frontmatter@RRAPformat}
  {}{}
}%
\makeatother

\begin{document}

\preprint{AIP/123-QED}

\title[Josephson coupling in Lanthanum-based cuprates superlattices]{Josephson coupling in Lanthanum-based cuprates superlattices}
\author{H. G. Ahmad}%
\email{halimagiovanna.ahmad@unina.it}
\affiliation{Department of Physics "Ettore Pancini", University Federico II, Naples, 80126, Italy.
}%
\author{D. Massarottti}%
 \affiliation{Department of Electrical Engineering and Information Technology, University Federico II, Naples, 80126, Italy.}
\author{F. Tafuri}%
\affiliation{Department of Physics "Ettore Pancini", University Federico II, Naples, 80126, Italy.
}%
\author{G. Logvenov}%
 \affiliation{Max Planck Institute for Solid State Research, Heisenbergstraße 1, 70569 Stuttgart, Germany.}
 \author{A. Bianconi}
 \affiliation{RICMASS Rome International Centre Materials Science Superstripes Via dei Sabelli 119A, 00185 Rome, Italy.}
\author{G. Campi}%
\affiliation{Institute of Crystallography, National Research Council of Italy, Via Salaria Km 29.300, Monterotondo, 00015 Rome, Italy.}


\begin{abstract}

In most anisotropic compounds such as bismuth-based layered cuprate perovskites, the supercurrent across the blocking layer is of Josephson type, and a single crystal forms a natural stack of Josephson junctions. Here, we report on the evidence of Josephson-like transport in an artificial cuprate superlattice composed of 10 LaSrCuO-LaCuO repeats, creating a superlattice of junctions, where LCO is a superconducting Mott insulator and LSCO an overdoped metal, respectively.  The superlattice has been designed with a long period $d=L+W=5.28\;nm$, with $L$ and $W$ the thickness of LCO and LSCO units, respectively, and is in the underdoped regime with an average doping level $<\delta>=0.11$. Quantum-size effects and Rashba spin-orbit coupling are controlled by $L/d = 0.75$, with a quasi-2D superconducting transition temperature of $41\;K$ and a c-axis coherence length of about $1.5\;nm$. Measurements at very low temperatures show evidence of Josephson phase dynamics consistent with very low Josephson coupling and a phase diffusion regime, thus explaining why Josephson coupling in LSCO superlattices has been so elusive. The tuning of LSCO superlattices in the Josephson regime enriches the phase diagram of HTS.
\end{abstract}
\maketitle

\section{\label{sec:level1}Introduction}

Current research on quantum materials for quantum technology focuses on quantum Nanostructured Materials (NsM) with novel functionalities~\cite{manna2023}, which can be tuned by controlling the quantum size effects of their nano-objects, offering unique opportunities to bridge quantum science from the micro to the macro world. In this context, high-$T_C$ cuprate superconductors are a special class and can now be engineered using the quantum design of nanoscale artificial high-Tc superlattices (AHTS) of quantum wells~\cite{logvenov2023, valletta2024, mazziotti2022, campi2024}, capturing key features of naturally chemically doped cuprates~\cite{fratini2010, campi2021, campi2015, campi2013}. Josephson coupling in AHTS is believed to provide unique benchmarks for advances in the understanding of their physics~\cite{bozovic2003}.

\begin{figure}[t]
\includegraphics[width=0.75\textwidth]{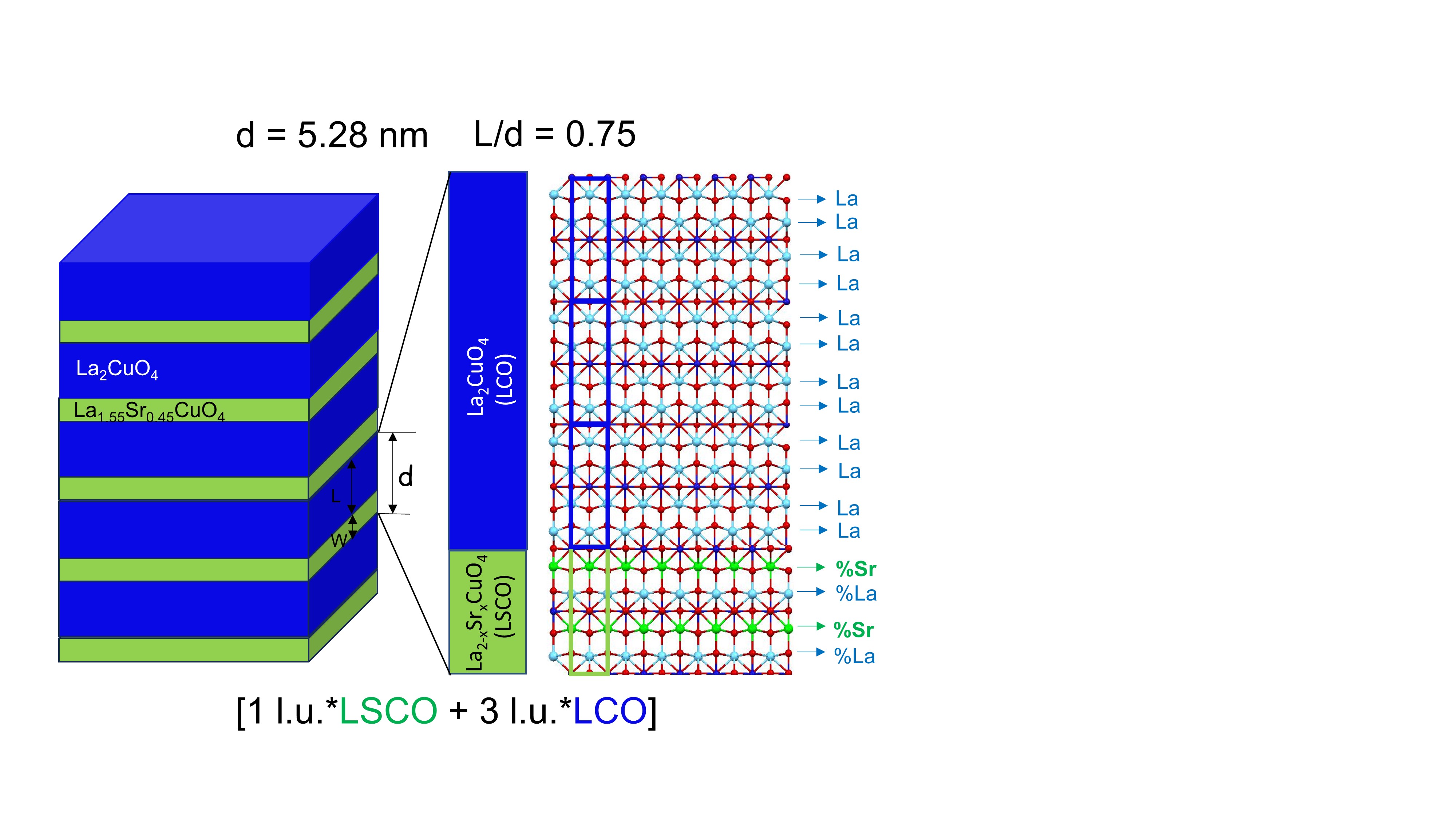}
\caption{\label{fig:1} Pictorial view of the nanoscale artificial high $T_C$ superlattice (AHTS) of quantum wells made of four monolayers ML ($L = 3.96\;nm$) of stoichiometric La$_2$CuO$_4$ (LCO), electronically doped by the interface space charge, and intercalated by normal metal units (LSCO) made of two monolayers ML ($W = 1.32\;nm$) of $La_{1.55}Sr_{0.45}CuO_4$ forming a superlattice with a period of $d = L+W =5.28\;nm$ and with conformational parameter $L/d = 0.75$.}
\end{figure}

We studied three-dimensional AHTS heterostructures composed of superconductor layers (S) of stoichiometric modulation-doped Mott insulator La$_2$CuO$_4$ (LCO) with thickness $L$, intercalated with potential barriers of normal metal (N) units made of chemically overdoped non-superconducting La$_{1.55}$S$_{0.45}$CuO$_4$ (LSCO). These structures have ten repeats of period $d = 5.28\;nm$. The N units act as charge reservoirs, transferring interface space charge into the superconducting Mott insulator units, forming a 3D superlattice of a 2D Mott insulator-metal interface (MIMI) with period $d$. A schematic of our 3D superlattice is illustrated in Figure~\ref{fig:1}. 

Using a four-point van der Pauw configuration, we measured the in-plane conductivity for samples with $L/d$ ratios from $0$ to $1$ and periods $d$ from $2.97$ to $5.28\;nm$~\cite{campi2024}. The results show the superconducting dome predicted by the BPV theory, where the critical temperature, $T_C$, is a function of the geometrical parameter $L/d$, peaking at the magic ratio $L/d = 2/3$. At this ratio, the theory predicts that the Fano-Feshbach resonance enhances the superconducting critical temperature~\cite{valletta2024}. This work demonstrates that AHTS can be designed such as to exhibit Josephson coupling along the c-axis, but so far never been observed.   

Since the discovery of high-$T_C$ superconductors (HTS), numerous theoretical analyses and experimental attempts have been made regarding High-$T_C$ superconducting Josephson Junctions (HTS JJs)~\cite{Schrieffer2008, bozovic2003, bozovic2004, tafuri2019, golubov2004, fischer2007, yokoyama2007, bergeal2005}. The unconventional and controversial properties of HTS JJs have made it a rich topic of study over three decades, highlighting it as a relevant challenge in condensed matter physics and quantum materials science. Josephson coupling has been measured in various HTS JJs~\cite{Tafuri2005}. Due to the difficulty in growing high-quality artificial barriers, most of the successful HTS JJs have employed the weak coupling between HTS grains~\cite{Hilgenkamp2002,Tafuri2005}, i.\,e. grain boundary JJs (GBJJ), which mostly consist of bicrystal~\cite{Chaudhari1988,Dimos1990}, step-edge~\cite{Luine1992,Yi1995} and biepitaxial JJs~\cite{Char1991,Tafuri1999,Lombardi2002,Bauch2005}. The natural stack of Josephson junctions in anisotropic cuprates is another distinctive class of Josephson devices, where the Josephson effect is used to probe the layered superconductor “from inside”~\cite{Kleiner2019}. Both GBs and c-axis coupling in intrinsic JJs are peculiar to HTS. However, there is no generally accepted understanding of the high-$T_C$ Josephson effect, and a high-quality device based on high-$T_C$ JJ of stacked SNS or SIS type has not been realized. 
This is due to challenges such as controlling the heterogeneous nanoscale structure, the nanoscale coherence length in the c-axis direction, and the difficulty in selecting and growing suitable barriers~\cite{bozovic2003, bozovic2004, shi2013, bollinger2012, stornaiuolo2019, zhou2019}. On the other hand, such JJ geometry would benefit superconducting electronics applications, including digital and quantum superconducting circuit fabrication, which require significant multi-layer integrated technology~\cite{Simon1990}. 

Recently, LSCO heterostructures and superlattices have been synthesized to study HTS c-axis JJs and interfacial superconductivity. LSCO-LCO-LSCO trilayer junctions, where the LCO barrier was annealed to form SIS, SNS, and SS'S types of junctions, have not yet resolved the open questions~\cite{xu2023, bollinger2023}. In this study, we report on the evidence of Josephson-like transport JJ in an artificial HTS superlattice composed of $10$ LSCO-LCO repeats, forming a superlattice of junctions. The nanoscale S units are stoichiometric La$_2$CuO$_4$ layers, modulation-doped by interface space charge with a c-axis coherence length of approximately $1.5\;nm$, while the N units are metallic overdoped La$_{1.55}$Sr$_{0.45}$CuO$_4$ thin layers. 

In Sec.~\ref{Methods}, we will report on the methods for the synthesis of the AHTS analyzed in this work. Then, in Sec.~\ref{results} we will focus on the experimental transport investigation down to liquid helium temperature of in-plane superconductivity, and on the study of Josephson transport along the c-axis direction of AHTS through a systematic characterization of the current-voltage (I-V) characteristics and conductance spectra down to dilution temperatures.

\section{Methods}
\label{Methods}

With the final goal of investigating the out-of-plane conductivity along the c-axis of AHTS superlattices, we have selected a sample fabricated by using a Molecular Beam Epitaxy (MBE) synthesis. We have chosen to focus on a sample with a very long period, $d = 5.28\;nm$, to hinder out-of-plane conductivity. It is known that maximum $T_C$ occurs at $L/d = 2/3$ ~\cite{logvenov2023, valletta2024, mazziotti2022, campi2024} where the average doping is $0.15$ as in natural HTS cuprates. The chosen sample has $L/d$ ratio of $0.75$ and an average doping $0.11$ in the underdoped regime. While in natural cuprates the superconducting temperature at this doping level is lower than the maximum $T_C$ because of competition with Charge Density Waves (CDW)~\cite{Miao2021}, in our AHTS the superconducting dome becomes larger and the superconducting temperature is close to the maximum $T_C$, indicating that the competition with CDW is quenched. 
\begin{figure}[t]
\includegraphics[width=0.5\textwidth]{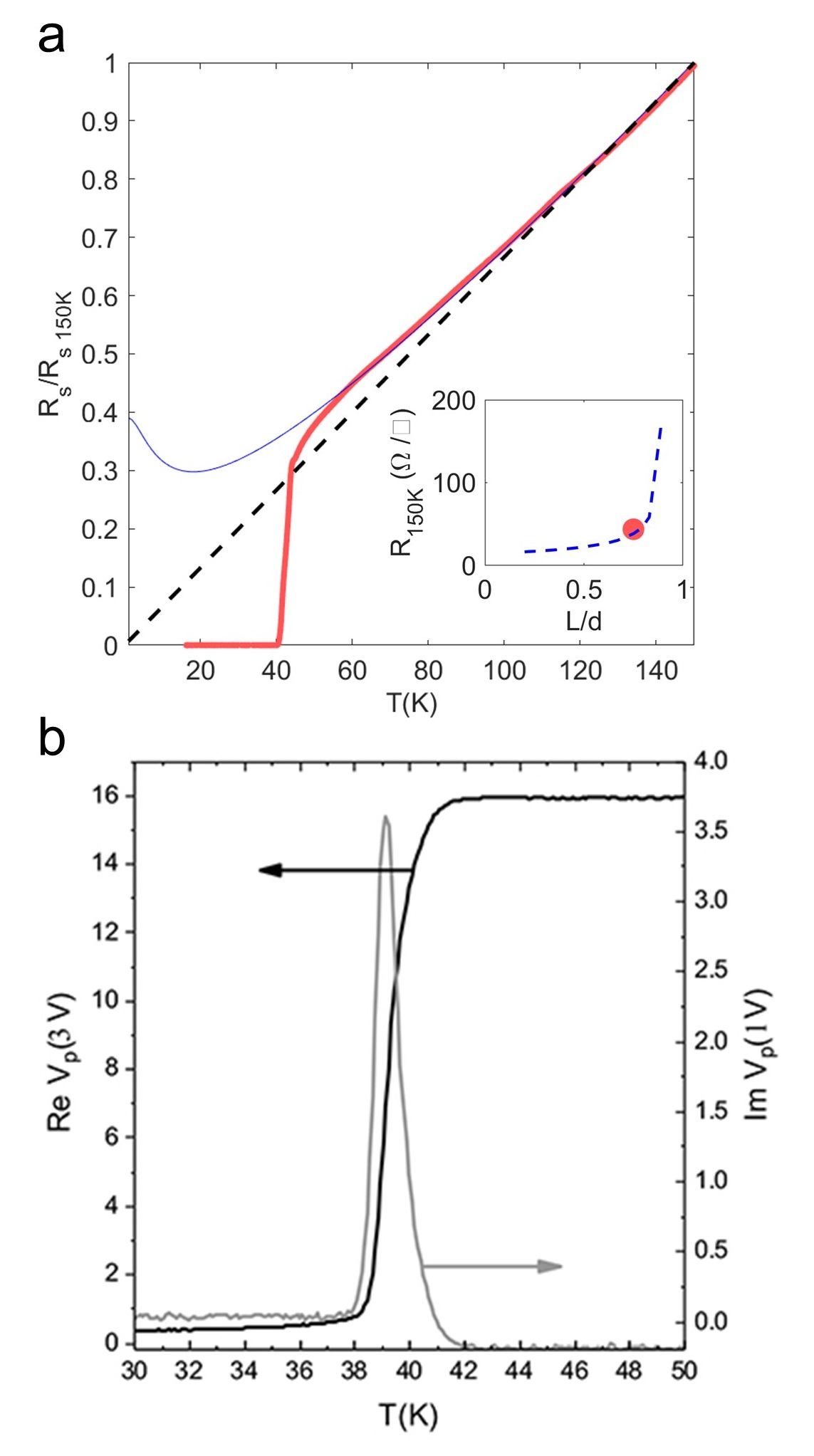}
\caption{\label{fig:2} (a) Sheet resistance  (red line) in $R_s(150\;K)$ units; the linear regime is represented by the black dashed line. The blue line represents the modeled line through Eq. \ref{eq:Kondo}. In the inset is reported the sheet resistance measured at $150\;K$ in several samples with different $L/d$ values (dashed blue line). The red full circle refers to the sheet resistance measured at $150\;K$ in the sample studied in this work.  (b) Real and Imaginary Components of Resistivity Measurement.}
\end{figure}

The MBE synthesis of AHTS is based on the growth of superlattices made of nanoscale non superconducting normal metal La$_{1.55}$Sr$_{0.45}$CuO$_4$ (LSCO) layers intercalated by superconducting space-charge La$_2$CuO$_4$ Mott insulator (LCO) layers. The oxide heterostructures at atomic limit were synthesized using an ozone-assisted MBE method (DCA Instruments Oy, Turku, Finland) on LaSrAlO$_4$ (001) substrates with a small $+1.4 \%$ compressive strain for La$_2$CuO$_4$ on LaSrAlO$_4$. The superlattice atomic growth was monitored in-situ by reflection high-energy electron diffraction (RHEED). The sequential deposition of single atomic layers was performed using impinging atoms of minimal kinetic energy of approximately $100\;meV$. The substrate temperature ($T_s$), measured by a radiation pyrometer, was $650$ °C, and the chamber pressure was approximately $1.5\times10^{-5}\;Torr$, consisting of mixed ozone, atomic, and molecular oxygen. After the synthesis at high temperature, the samples were cooled down to low temperature avoiding trapping of oxygen interstitials in stoichiometric La$_2$CuO$_4$ Mott insulator.

\section{Results and discussion}
\label{results}

\subsection{In-plane superconductivity}

The temperature dependence of the sheet resistance in the ab planes of AHTS was measured using a four-point van der Pauw configuration~\cite{Pauw1958} with alternating DC currents of $\pm10\;\mu A$, over a temperature range from room temperature to liquid helium. We have used a motorized custom-made dipstick in a transport helium dewar maintaining a temperature rate of less than $0.1\;K/s$. The measured sheet resistance corresponds to the red line in Figure~\ref{fig:2} (a). According to recent work~\cite{campi2024}, the modeling curve (blue line) has been obtained by the generalized Kondo equation for Mott Insulator–Metal Interfaces (MIMI)~\cite{costi1994, helmes2008}. The experimental $R(T)$ curve is modeled by combining Planckian T-linear resistance and the universal resistance function for the Kondo proximity effect at Mott-Insulator-Metal interfaces. The formula used is:
\begin{equation}
\textstyle\frac{R(T)}{R(T=150K)} = r_0 + \frac{T}{150} + AT^2 + BT^5 + \frac{R_{0K}}{\left[1 + (2^{1/s} - 1) \left(\frac{T}{T_K}\right)^2\right]^s}. \label{eq:Kondo} 
\end{equation}
This model applies to experimental $R(T)$ values measured at temperatures above $50\;K$, in the normal phase. The Kondo temperature $T_K$ is the characteristic temperature below which the localized spin is screened by conduction electrons. $R_{0K}$ represents the amplitude of the Kondo-like contribution, while $r_0$ is the residual contribution at $T=0$, indicating the baseline resistivity without Kondo physics. The terms $AT^2$ and $BT^5$ account for electron-electron and phonon scattering, respectively, and are non-Kondo contributions. The exponent $s$ is a material-dependent parameter depending on the Metal-Mott Insulator Interfaces and here is found to be around $0.12$. This distinctive contribution to resistance can fit experimental $R(T)$ down to about $T_c$. When further lowering the temperature, in absence of the superconducting transition, this term would determine an increase of the resistance. A least squares fit algorithm was used to extract parameters $A$, $B$, $R_{0K}$ and $T_K$. The Kondo temperature, obtained from the fitting, is $T_K= 15.9\;K$, smaller than $T_C$ but larger than the $T_K = 4.5 \pm 1\;K$ at optimum superconductivity achieved at $L = 2/3$. In sheet resistance measurements at $150\;K$ in several samples with different values of $L/d$,~\cite{campi2024} an exponential increase was observed for $L/d > 2/3$, indicating a metal-insulator transition triggered by the geometric conformational parameter, as shown in the inset of Figure~\ref{fig:2} (a). Here the red full circle represents the sample studied in this work. In this sample the long period and the low doping level are expected to increase the sheet resistance at $150\;K$ in the normal phase. The sample shows a superconducting transition of $T_C=41\;K$ and at $39\;K$ by measuring mutual inductance (Figure~\ref{fig:2} (b)).

\subsection{Out-of-plane low temperature transport}

Transport experiments along the c-axis have been run at dilution temperatures using an Oxford Instruments Triton400 dry dilution fridge, anchoring the sample to a copper holder mounted at its coldest stage, i.\,e. the mixing chamber (MXC) plate (nominal base temperature of $\sim10\;mK$). The cryogenic system is equipped with filtered current and voltage cryogenic DC lines, specially engineered to run low-noise transport experiments on Josephson-based devices. This allows us to measure the voltage response of current-biased devices down to the nanoampere regime~\cite{Vettoliere2022,Ahmad2024}.  Details on the experimental setup can be found in the Supplementary Material in Ref.~\cite{Ahmad2024} and Ref.~\cite{Satariano2024}. 

\begin{figure}[t]
\includegraphics[width=0.45\textwidth]{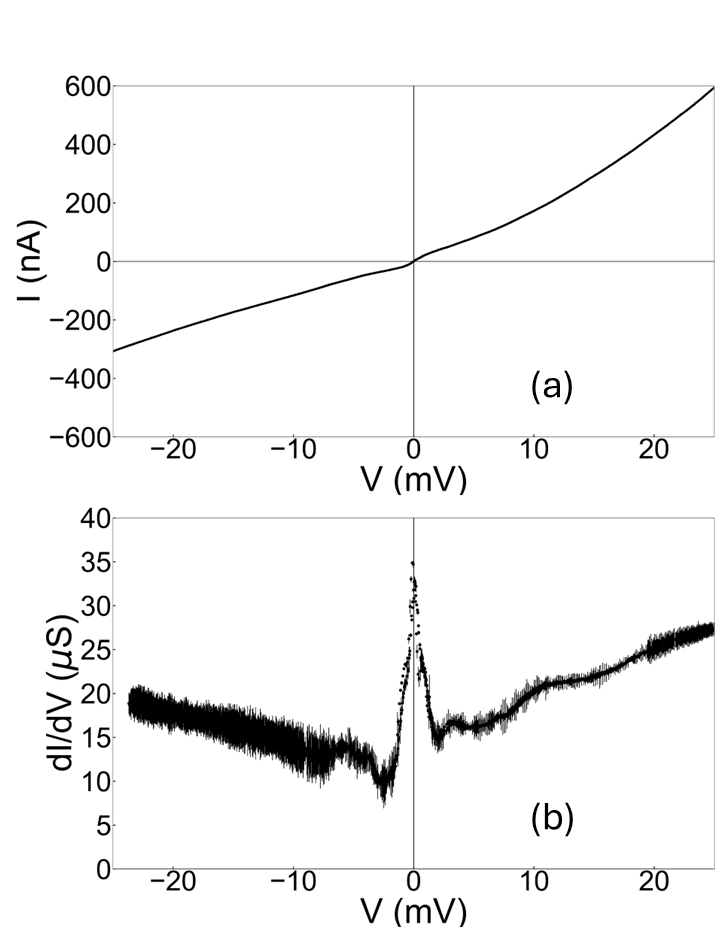}
\caption{\label{fig:long_range} (a) Current-voltage characteristic I-V at $12\;mK$ in the presence of a quasi-dc triangular current bias (i.e., for a current ramp frequency of $1.123\;Hz$), with a peak-to-peak bias current of $1.2\;\mu A$). The current and voltage errors are maximum errors of $1$ and $2\%$, respectively. In (b), conductance spectrum $dI/dV$ at $12\;mK$, for a polarization current composed of a sinusoidal current excitation at $31.123\;Hz$, superimposed to a quasi-dc low-frequency triangular ramp of $1.123\;mHz$. The peak-to-peak amplitude of the low-frequency current polarization is the same as in a), while the high-frequency component is of the order of $5/1000$ of the low-frequency signal amplitude. The error on $dI/dV$ is of the order of $7\%$.}
\end{figure}

The device has been connected to the cryogenic and room-temperature electronics, in such a way that the current bias, provided by an arbitrary waveform generator at room temperature, flows along its c-axis. The voltage across the device has been first amplified at room temperature by a differential low-noise amplifier, and then readout with an oscilloscope~\cite{Ahmad2023}. An example of the current-voltage I-V characteristic at base temperature in a large range of current bias is reported in Figure~\ref{fig:long_range} (a).

The intrinsic complexity of HTS JJs requires great accuracy when establishing general features and self-consistency of the measured set of parameters. This strengthens the need for complementary types of measurements~\cite{tafuri2019} and motivates the parallel study of conductance spectra experimental curves, i.\,e. $dI/dV$ characteristics, both at base temperature and as a function of temperature. In the measurement of the dI/dV characteristics, we current-bias the device with a sum of two current ramp signals: one triangular current ramp at low frequency ($1.123\;mHz$) provided by the arbitrary waveform generator, and the other sinusoidal at high frequency ($31.123\;Hz$) and with a peak-to-peak amplitude of the order of $5/1000$ of the low-frequency signal, generated by a lock-in amplifier~\cite{Vettoliere2022}. The latter has been used also to read out the voltage response $dV$ of the device to this excitation $dI$. In Figure~\ref{fig:long_range} (b) we provide the AHTS $dI/dV$ characteristic at base temperature in a similar range of current bias as in panel (a). In Fig.~\ref{fig:long_range}, we observe a  remarkable asymmetric behavior, especially at high voltages, in the non-linear asymmetric current-voltage (AIV) curve with respect to inversion of voltage. A similar asymmetric behavior was observed in the $dV/dI$ characteristics of the very first cuprate-based JJs, which has been generically related to an overall Schottky nature of the barrier~\cite{Kleiner2019}. Therefore, the observed AIV curve could be assigned to the different top and bottom layer of the AHTS superlattice and to the asymmetry of the interface potential barrier between the doped Mott insulator LCO (S) and the normal metal LSCO (N).  This is related to the contact electric field controlling the Rashba Spin Orbit Coupling (SOC) at the interface between metal oxide layers~\cite{Gariglio2019,Chen2024}, which is a key ingredient in the material quantum design of the oxide superlattices. 

\begin{figure}[t]
\includegraphics[width=0.45\textwidth]{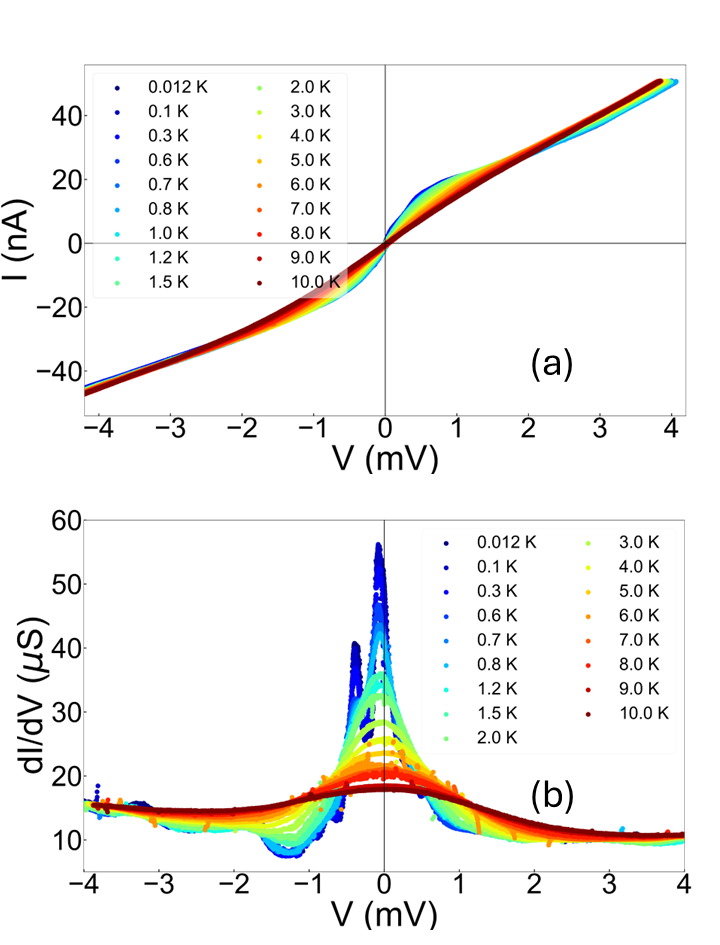}
\caption{\label{fig:temp} (a) Current-voltage characteristic I-V as a function of the temperature in the presence of a quasi-dc triangular current bias (i.e., for a current ramp frequency of $1.123\;Hz$, with a peak-to-peak bias current of $84.6\; nA$). The current and voltage errors are maximum errors of $1$ and $2\%$, respectively. In (b), conductance spectrum $dI/dV$ as a function of the temperature. The experimental conditions and errors are the same as the experimental data reported in Fig.~\ref{fig:long_range}.}
\end{figure}

Both I-V and $dI/dV$ characteristics have been measured at different temperatures in a shorter current-bias range, to focus on the superconducting branch and the superconducting to the resistive state switching of the device (Figure~\ref{fig:temp}). As reported in the data below, a weak Josephson coupling along c-axis appears at very low temperatures. As opposed to previous attempts on similar LaSCO-based systems~\cite{bozovic2003}, where no Josephson coupling was measured, in the present experiment it was crucial to reach very low temperatures and to apply very low bias currents. The critical current $I_c$, defined here as the point in which the I-V changes its derivative as a function of the bias current, is in fact of the order of $20\;nA$ at the lowest investigated temperature, and consistent with a low critical current density.
\begin{figure}[t]
\includegraphics[width=0.45\textwidth]{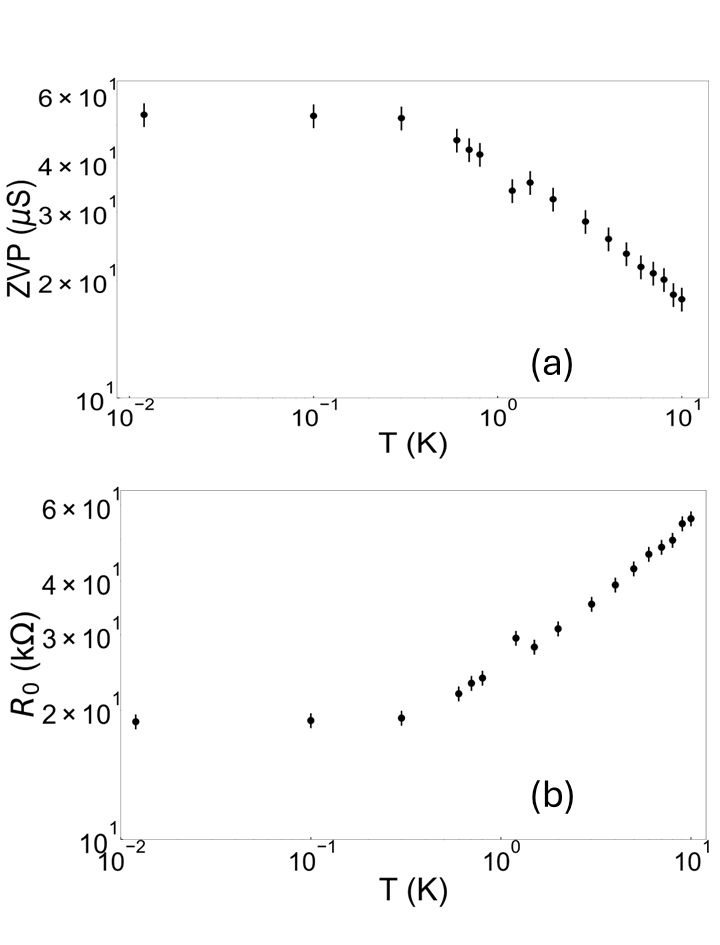}
\caption{\label{fig:ZVP} (a) The dependence of the zero-voltage-peak ($ZVP$) derived from Fig.~\ref{fig:temp} (b) as a function of the temperature $T$ in a logarithmic scale. In (b), the dependence of the resistance in the superconducting branch, $R_0$, as a function of $T$ in a logarithmic scale, derived from the I-V characteristics in Fig.\ref{fig:temp} (a).  The error on the resistance in panel (b) is given by propagation of voltages and current maximum errors and is $4\%$.}
\end{figure}
Switching from the superconducting to the resistive state does not occur abruptly as in more standard Josephson devices, showing a pronounced rounding of the I-V at the switching between the resistive to the superconducting state. The I-V characteristics of the device show a finite slope $R_0$ in the superconducting branch. The finite resistance measured in the superconducting branch also emerges in $dI/dV$ spectra as a peak corresponding to zero voltage, or Zero-Voltage-Peak ($ZVP$). 

Both the rounding of the I-V characteristics at the superconducting to resistive switching and the presence of a finite slope in the superconducting branch, as well as its thermal dependence, are clear markers and a direct confirm of the occurrence of phase-diffusive phenomena, rigorously controlled by the intrinsic features of the junction and its low critical current density~\cite{Iansiti1987,Kautz1990}. Phase-diffusion is an electrodynamical process for which the phase-particle of the JJ randomly escapes and retraps in the minima of the washboard potential energy landscape of the JJ, specifically when the Josephson energy $E_J=\hbar I_c/2e$ is $\ll k_BT$, or in the case of the competition between $E_J$ and the charging energy $E_c$ of the JJ~\cite{Iansiti1987,Martinis1989,Kautz1990,Vion1996}, and can be accounted for by modeling the ATHS JJ as an RLC parallel circuit with the introduction of electronic random Gaussian noise in the frame of an extended version of the RCSJ (Resistively and Capacitively Shunted Junction) model~\cite{Kautz1990}. In the frame of HTS JJs, phase-diffusion phenomena have been already demonstrated to occur in different regimes of energy scales and electrodynamics parameters~\cite{Warburton2004,Bae2009,Longobardi2012,Stornaiuolo2013}. Just because the measured $I_c$ for the AHTS devices analyzed in this work are of the order of a few tens of nanoamperes at dilution temperatures, the ratio $E_J/k_B T$ is already consistent with the presence of phase-diffusion phenomena. 

As the temperature increases, $E_J$ is outruled by the thermal energy $k_BT$, further lowering the resolution of the I-V characteristics. We discuss the behavior as a function of temperature of the transport by looking at the $ZVP$ and the $R_0$ in Figure~\ref{fig:ZVP} (a) and (b), respectively. The $ZVP$ follows an exponential decrease when increasing the temperature: this indicates that superconducting transport is suppressed when thermal energy becomes dominant. Moreover, no hysteresis has been observed in the I-V characteristics when applying a triangular current bias on the device. These two pieces of evidence are the fingerprints of typical overdamped JJs~\cite{Likharev1979}. 

We also point out here that the $ZVP$ has been identified as the maximum peak in Figure~\ref{fig:long_range} (b). However, one can notice that within the resolution of our experiment, two distinguishable peaks emerge in the $dI/dV$ characteristics, which tend to broaden when increasing the temperature, and eventually merge at around $0.8\;K$. At very low temperatures, one can notice also additional less-resolved minor peaks. The presence of multiple peaks in the $dI/dV$ may be related to multiple switching in the I-V characteristics. The latter effect remarkably reminds us of the superconducting transport in intrinsic stacked HTS JJs~\cite{Kleiner1992,Latyshev1999}, where each of the planes along the c-axis plays the role of a JJ in series, and multiple switchings relate to the transition of each plane of the device. This not only is consistent with the periodic structure of the AHTS, but is also a starting point for further discussion on the $R_0$ thermal behavior.

In principle, if each plane of the superlattice along the c-axis shows the Josephson effect, one should see as many peaks as many layers the device has. Nevertheless, the ATHS critical current value related to the maximum $ZVP$ is three to six orders of magnitudes lower than the one typically measured in long weak links~\cite{Likharev1979,tafuri2019} or similar stacked HTc intrinsic JJs~\cite{Kleiner1992,Chana2000,Yurgens2000}, thus making hard to resolve other critical current steps $I_{c,n}$ of the $n$-th layer. As a consequence, when current-biasing the device with $I>I_{c,n}$ of the $n$-th layer, a finite resistance sums up in a series-like fashion. Since we can not exclude that other resistive contributions occur due to spurious ohmic contacts in the device, only a qualitative description can be given at this stage by looking at the thermal behavior of $R_0$. In figure~\ref{fig:ZVP} (b), while at base temperature a finite resistance is measured, by increasing the temperature the superconducting branch slope increases compared to this starting value, up to the point we just observe a linear trend of the I-V characteristic at $10\;K$. 

\section{Conclusion}

In this work we present compelling evidence for intrinsic overdamped JJs in an artificial high-$T_C$ superlattice composed of $10$ LCO/LSCO repeats with a $5.28\;nm$ period, forming a superlattice of junctions. The nanoscale S units are stoichiometric La$_2$CuO$_2$ layers, $3.9\;nm$ thick. The superconducting c-axis coherence length is expected to be approximately $1.5\;nm$, similar to the thickness of the N units. The metallic overdoped LaSrCuO thin layers, acting as potential barriers, consist of a single La$_2$CuO$_2$ unit cell of $1.32\;nm$. This very first evidence of the Josephson effect in lanthanum-based cuprates, occurring at very low temperatures, is consistent with phase diffusion dynamics, and with very low Josephson coupling. We can speculate that the variation of the diffusion regime along the c-axis at $T<10\;K$ could be related to the Kondo temperature found in the sheet resistance of the normal phase.  These findings open new perspectives for extending the applications of intrinsic Josephson junction stacks observed in Bi$_2$Sr$_2$CaCu$_2$O$_8$ (BSCCO or Bi2212) crystals~\cite{itoh1997, ji2014, lang2023, charikova2024} and in artificial devices using Josephson junctions made of twisted ultrathin Bi$_2$Sr$_2$CaCu$_2$O$_8+y$ flakes~\cite{zhu2021, poccia2020, martini2023, brosco2024}, to the synthesis of novel quantum devices using superconducting nanoscale units, and the Josephson effect in Lanthanum-based cuprates superlattices. The overdamped nature of the devices analyzed here favors applications in SQUID technology. Further advanced engineering of the superlattice in terms of the junction geometry and parameters can better define the application domain for these devices.

\begin{acknowledgments}

The work was supported by Superstripes onlus, the Pathfinder EIC 2023 project "FERROMON-Ferrotransmons and Ferrogatemons for Scalable Superconducting Quantum Computers" (Grant Agreement ID: 101115548), the PNRR MUR project PE0000023-NQSTI and the PNRR MUR project CN 00000013-ICSC. H.G.A., D.M. and F.T. thank SUPERQUMAP project (COST Action CA21144) and Dr. Nicola Poccia for his invaluable insights on the work.

\end{acknowledgments}

\section*{Data Availability Statement}

The data that support the findings of this study are available from the corresponding author upon reasonable request.

\section*{Conflict of Interest statement}

The authors have no conflicts to disclose.

\nocite{*}

	\providecommand{\noopsort}[1]{}\providecommand{\singleletter}[1]{#1}%
	

\end{document}